\documentclass[fleqn,usenatbib]{mnras}

\usepackage{graphicx}	
\usepackage{amsmath}	
\usepackage{amssymb}	
\usepackage[dvipsnames]{xcolor} 
\usepackage{ulem}

\newcommand{\nenya}{{\tt nenya}}
\newcommand{\vilya}{{\tt vilya}}
\newcommand{\narya}{{\tt narya}}

\newcommand{\paf}{``Paired-\&-Fixed''}

\newcommand{\Msun}{ h^{-1}{\rm M_{ \odot}}}
\newcommand{\hMpc}{ h^{-1}{\rm Mpc}}
\newcommand{\hkpc}{ h^{-1}{\rm kpc}}
\newcommand{\ihMpc}{ h\,{\rm Mpc}^{-1}}

\title[The BACCO simulations project]{The BACCO Simulation Project: Exploiting the full power of large-scale structure for cosmology}

\author[Angulo et al.]{
Raul E. Angulo,$^{1,2}$\thanks{E-mail:reangulo@dipc.org}
Matteo Zennaro$^{1}$,
Sergio Contreras$^{1}$,
Giovanni Aric\`o,$^{1}$, \and 
Marcos Pellejero-Iba\~nez$^{1}$,
\& Jens St\"ucker$^{1}$.
\\
$^{1}$Donostia International Physics Center (DIPC), Paseo Manuel de Lardizabal, 4, 20018, Donostia-San Sebasti\'an, Guipuzkoa, Spain.\\
$^{2}$IKERBASQUE, Basque Foundation for Science, 48013, Bilbao, Spain.\\
}

\date{Accepted XXX. Received YYY; in original form ZZZ}

\pubyear{2019}

\begin{document}
\label{firstpage}
\pagerange{\pageref{firstpage}--\pageref{lastpage}}
\maketitle

\begin{abstract}
We present the BACCO project, a simulation framework specially designed to provide highly-accurate predictions for the distribution of mass, galaxies, and gas as a function of cosmological parameters. In this paper, we describe our main suite of {\bf gravity-only} simulations ($L\sim2\,$Gpc and $4320^3$ particles) and present various validation tests. Using a cosmology-rescaling technique, we predict the nonlinear mass power spectrum over the redshift range $0 < z < 1.5$ and over scales $10^{-2} < k/(\ihMpc) < 5$ for $800$ points in an $8$-dimensional cosmological parameter space. For an efficient interpolation of the results, we build an emulator and compare its predictions against several widely-used methods. Over the whole range of scales considered, we expect our predictions to be accurate at the $2\%$ level for parameters in the minimal $\Lambda$CDM model and to $3\%$ when extended to dynamical dark energy and massive neutrinos. We make our emulator publicly available under \url{http://www.dipc.org/bacco}
\end{abstract}




\section{Introduction}

Over the last two decades, our understanding of the Universe has grown tremendously: the accelerated expansion of the Universe and the existence of dark matter are firmly supported; there are strong constraints on the micro-physical properties of dark matter and neutrinos; and the statistics of the primordial cosmic fluctuations are measured with increasing precision \citep[e.g.][]{Alam:2017,Planck2018,Gilman:2020}. 

Despite the progress, there are several tensions among current data when interpreted within the context of the $\Lambda$CDM model. For instance, the value of the Hubble parameter inferred from local supernovae is significantly larger than that inferred from the analysis of the CMB \citep[e.g.][]{Riess:2019,Freedman:2019}. Another example is that the amplitude of fluctuations, $S_8 = \sqrt{\Omega_m/0.3}\,\sigma_8$, as determined from low-redshift lensing measurements, appears smaller than that inferred from CMB \citep[e.g.][]{Asgari:2020}.  

In the current era of precision cosmology, the large amount of available large-scale structure (LSS) data promises accurate measurements of cosmological parameters with small systematic errors. These future measurements could shed light on the aforementioned tensions, confirming or ruling out the $\Lambda$CDM paradigm \citep[e.g.][]{Weinberg2013}. Many observational campaigns that seek to obtain this data are under construction or with an imminent start (e.g. Euclid\footnote{\url{https://sci.esa.int/web/euclid}}, DESI\footnote{\url{https://www.desi.lbl.gov/}}, J-PAS\footnote{\url{http://j-pas.org}}).

To fully exploit the upcoming LSS data and obtain these cosmological constraints, extremely accurate theoretical models are required. This is an active area of research with different approaches being adopted in the literature. On the one hand, recent advances in perturbation theory have increased considerably the range of scales that can be treated analytically \citep[e.g.][]{Desjacques:2018}. These scales are, however, still in the quasi-linear regime. On the other hand, cosmological numerical simulations 
have also steadily increased their robustness and accuracy \citep[e.g.][]{Kuhlen:2012}. Currently, simulations are the most accurate method to model smaller and non-linear scales (where, in principle, much more additional constraining power resides). 

Traditionally, numerical simulations were very expensive computationally and suffered from large cosmic-variance errors, thus they were only used to calibrate fitting functions or combined with perturbation theory to provide predictions for nonlinear structure as a function of cosmology \cite[e.g.][]{Smith:2003,Takahashi:2012}. This has changed recently, as the available computational power keeps increasing and algorithms become more efficient, it is now possible carrying out large suites of simulations spanning different cosmological models. The so-called emulators have become more popular, which interpolate simulation results to provide predictions in the nonlinear regime and for biased tracers \citep[e.g.][]{Heitmann:2014,Nishimichi:2018,Liu:2018,DeRose:2019,Giblin:2019,Wibking:2019}.

To keep the computational cost affordable, emulators are typically restricted to a small region in parameter space which is sampled with a small number of simulations ($\sim50-100$). Additionally, each individual simulation is low resolution or simulates a relatively small cosmic volume. This limits their usability in actual data analyses and/or add a significant source of uncertainty.

Here we take advantage of several recent advances to solve these limitations. First, we employ a very efficient $N$-body code to carry out a suite of 6 large-volume high-resolution simulations, which allows to resolve all halos and subhalos with mass $>5\times 10^{10}\Msun$, together with their merger histories. Second, we employ initial conditions with suppressed variance, which allows to predict robustly even scales comparable to our simulated volume. Third, these simulations are combined with cosmology-rescaling algorithms, so that predictions can be obtained for any arbitrary set of cosmological parameters. As a result, this approach allows us to make highly accurate predictions for the large-scale phase-space structure of dark matter, galaxies, and baryons.

Our approach has many advantages over others in the literature. Firstly, we can predict the matter distribution over a broad range of scales, with high force accuracy and over a large cosmic volume. This allows for detailed modellings of the distribution of gas and the impact of ``baryonic effects'' \citep{vanDaalen:2011,Schneider2016,Chisari:2018,Chisari:2019,Arico:2020,vanDaalen:2020}. We can also resolve collapsed dark matter structures and their formation history, which enables sophisticated modelling of the galaxies that they are expected to host \citep[e.g.][]{Henriques:2020,Moster:2018,ChavesMontero:2016}. In addition, the cosmological parameter space is large and densely sampled, so that emulator uncertainties are kept under a desired level. Finally, the parameter space includes non-standard $\Lambda$CDM parameters, dynamical dark energy and massive neutrinos.

As an initial application of our framework, we have used our suite of specially-designed simulations to predict the nonlinear mass power spectrum over the range $0<z<1.5$ for $800$ different cosmologies within an 8-dimensional parameter space defined by a $\sim10\sigma$ volume around Planck's best-fit values. From these, we construct and present an emulator so that these predictions are easily accessible to other researchers. Overall, we reach a few percent accuracy over the whole range of parameters considered.  

Our paper is structured as follows. Section \ref{sec:methods} is devoted to presenting our numerical simulations and to the description and validation of numerical methods. In Section \ref{sec:emulator} we describe the construction of an emulator for the nonlinear power spectrum. 
We highlight Section \ref{sec:emu_params} where we discuss our strategy for selecting training cosmologies, Section \ref{sec:emu_pk} which discusses our power spectra measurements, and Section \ref{sec:emu_comparison} that compares our predictions to other approaches. We summarise our results and discuss the implications of our work in Section \ref{sec:summary}.

\section{Numerical methods} \label{sec:methods}

\subsection{The BACCO Simulations} \label{sec:bacco_sims}

\begin{figure*}
\includegraphics[width=0.9\textwidth]{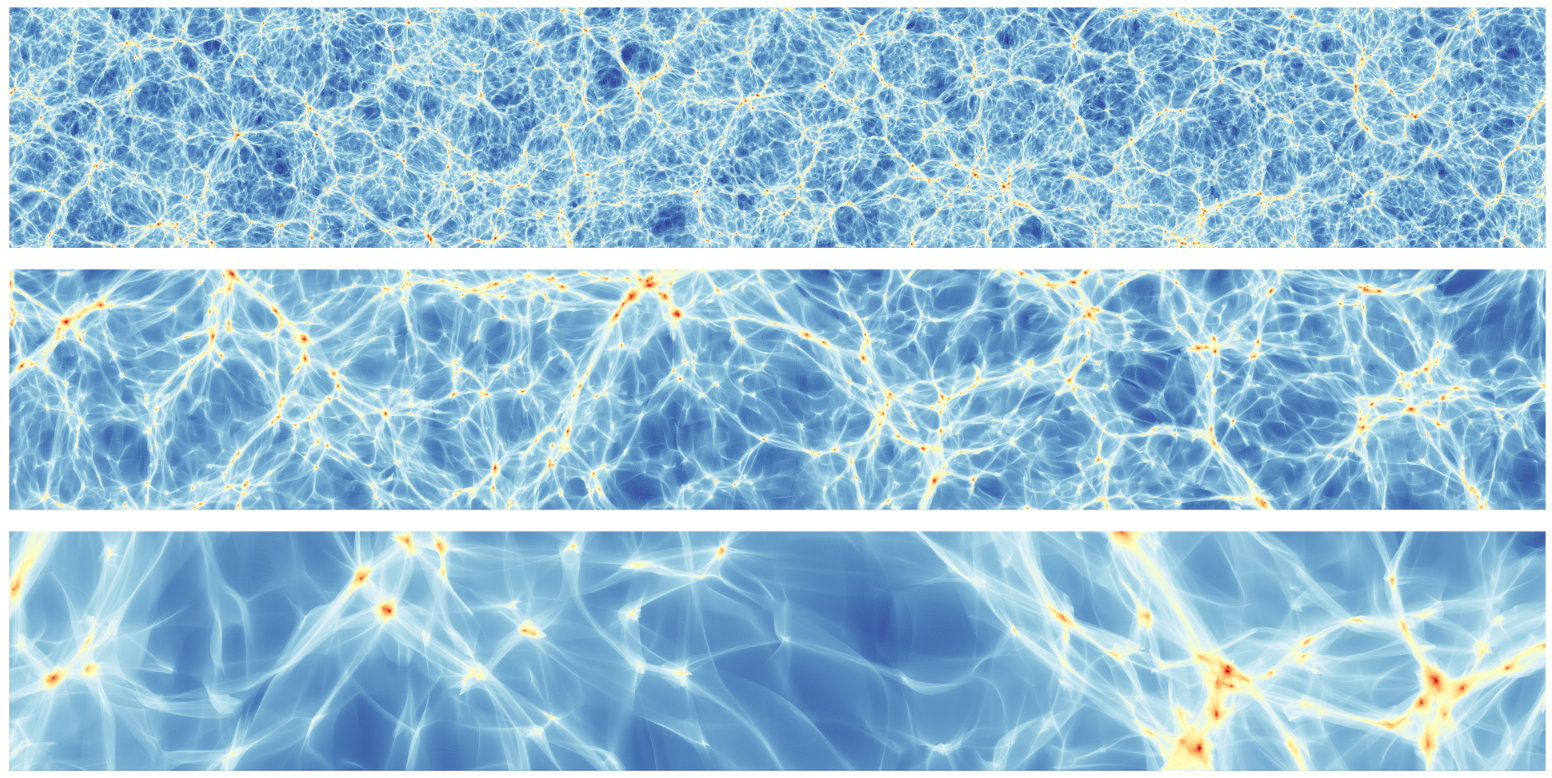}
\caption{The projected mass density field in \nenya, one of our six BACCO simulations, at $z=0$. Each image corresponds to a $25~\hMpc$ deep projection employing a tri-cubic Lagrangian interpolation method. Top, middle, and bottom panels progressively zoom into a $1440\,\hMpc$, $360\,\hMpc$, and $90\,\hMpc$-wide region of the simulation.    
\label{fig:density_field}}
\end{figure*}

The ``BACCO simulations" is a suite of 6 $N$-body simulations that follow the nonlinear growth of dark matter structure within a cubic region of L=$1440 \hMpc$ on a side. These calculations are performed for three different sets of cosmological parameters with two realizations each. The matter distribution is represented by $4320^3$ ($\sim 80$ billion) particles.  

The three cosmologies adopted by our BACCO simulations are provided in Table \ref{tab:parameters}. We dub these cosmologies \narya, \nenya, and \vilya \footnote{\narya, \vilya, and \nenya\ are the most powerful rings after Sauron's ``One Ring'' in ``The Lord of the Rings'' mythology.}. These sets are inconsistent with the latest observational constraints, but they were intentionally chosen to so that they can be efficiently combined with cosmology-rescaling algorithms \citep{AnguloWhite:2010}. Specifically, \cite{Contreras:2020} showed that these 3 cosmologies are optimal, in terms of accuracy and computational cost, to cover all the cosmologies within a region of $\sim10\sigma$ around the best values found by a recent analysis of the Planck satellite \citep{Planck2013}. 

For \narya\ and \vilya\ we stored $50$ snapshots, equally log-spaced in expansion factor, $a$, and adopt a Plummer-equivalent softening length of $\epsilon = 6.7\hkpc$. As pointed out by \cite{Contreras:2020}, most of the cosmological parameter volume is covered by rescaling \nenya, thus we have increased the force resolution and frequency of its outputs to $\epsilon = 5\hkpc$ and $100$ snapshots, respectively. All simulations were started at $a=0.02$ using 2nd-order Lagrangian Perturbation theory, and were evolved up to $a=1.25$ so that they can be accurately scaled to cosmologies with large amplitude of fluctuations.

For each of the three cosmologies, we carry out two realizations with an initial mode amplitude fixed to the ensemble mean, and opposite Fourier phases \citep{AnguloPontzen:2016, Pontzen:2016}. These \paf\, initial conditions allow for a significant reduction of cosmic variance in the resulting power spectrum in the linear and quasi-linear scales (which are the most affected by cosmic variance), as it has been tested extensively in recent works \citep{Chuang:2019,Villaescusa-Navarro:2018,Klypin:2020}. 

The BACCO simulations were carried out in the Summer of 2019 at Marenostrum-IV at the Barcelona Supercomputing Center (BSC) in Spain. We ran our $N$-body code in a hybrid distributed/shared memory setup employing $8192$ cores using $4096$ MPI tasks. The run time was 7.2 million CPU hours, equivalent of $35$ days of wall-clock time. The total storage required for all data products is about $80$ TB.  

To compare against recent emulator projects, we notice these simulations have $10$ times better mass resolution, $4$ times better force resolution, and $2\times3$ the volume of those used by the AEMULUS project \citep{DeRose:2019}; 50 times the volume and 3 times better mass resolution than MassiveNuS \citep{Liu:2018}; $6$ times better mass resolution, 50\% larger volume, and twice the spatial resolution than those in the EUCLID emulator project \citep{Knabenhans:2019}; $10$ times more particles and slightly better force resolution than the runs of the DarkEmulator \citep{Nishimichi:2018}; $2$ times better mass resolution and similar force resolution and volume to the simulations in the {\tt Mira-Titan} Universe project \citep{Heitmann:2014}. All this, of course, is mostly a result of our project being able to focus computational resources on only 6 simulations, but that is precisely the advantage of the cosmology rescaling approach.

In Fig.~\ref{fig:density_field} we display the projected matter density field of one of our simulations, \nenya\, at $z=0$. Each panel shows a region of the simulated volume, the top panel a region $1440\,\hMpc$ wide -- the full simulation side-length --, whereas middle and bottom panels zoom in regions $960$ and $360\,\hMpc$ wide, respectively. We display the density field as estimated via a tri-cubic Lagrangian tessellation using only $1080^3$ particles. Note that thanks to this interpolation, no particle discreteness is visible and filaments and voids become easily distinguishable.  

\begin{table*}
\caption{
The cosmological parameters of the three ``BACCO'' simulations presented in this paper: Vilya, Nenya and Narya}
\begin{center}
 \begin{tabular}{c c c c c c c c c c c c c}
 \hline
 & $\rm \sigma_8$ & $\rm \Omega_{\rm m}$ & $\rm \Omega_b$ & $\rm n_s$ & $\rm h$ & $M_{\nu}$ & $w_0$ & $w_a$ & $L$ [$\hMpc$] & $m_p$ [$\Msun$]& $\epsilon$ [$\hkpc$] &$N_p$\\ 
 \hline
 Vilya & 0.9 & 0.27  & 0.06 &  0.92 & 0.65 & 0.0 & -1.0 & 0.0  & 1440 & $2.77\times10^{9}$ & 6.7 & $4320^3$ \\ 
 Nenya & 0.9 &  0.315 & 0.05 &  1.01 & 0.60 & 0.0 & -1.0 & 0.0 & 1440 & $3.2\times10^{9}$ & 5 & $4320^3$\\ 
 Narya & 0.9 & 0.36  & 0.05 &  1.01 &  0.70 & 0.0 & -1.0 & 0.0 & 1440 & $3.7\times10^{9}$ & 6.7 & $4320^3$ \\ 
\end{tabular}
\end{center}
\label{tab:parameters}
\end{table*}

\subsection{The simulation code}
\label{sec:the_code}

The $N$-body code we employ is an updated version of {\tt L-Gadget3} \citep{Springel2005GADGET,Angulo:2012}. This code was originally developed for the Millennium-XXL project and was successfully executed with more than 10,000 CPUs employing 30Tb of RAM. Compared with previous versions of {\tt Gadget}, {\tt L-Gadget3} features an hybrid OpenMP/P-thread/MPI parallelisation strategy and improved domain decomposition algorithms. 

In addition to these improvements, our updated {\tt L-Gadget3} version stores all outputs in the HDF5 format, implements the possibility of simulating massive neutrinos via the linear response approach \citep{AliHaimoudBird:2013}, and features an improvement in the Tree-PM force split and in Kick-Drift operators. The output data structure is such to allow a straightforward reconstruction of the full phase-space distribution via tri-cubic Lagrangian interpolation \citep{Hahn:2016,Stucker:2020}.

The code carries out a large fraction of the required post-processing on the fly. Specifically, this includes the 2LPT initial conditions generator, group finding via Friends-of-Friends algorithms and an improved version of {\tt SUBFIND} \citep{Springel2001SUBFIND} that is also able to track tidally-disrupted structures, and a descendant finder and merger tree construction (c.f. \S\ref{sec:subfind}).

Thanks to the in-lining of these tools, it is not necessary to store the full particle load at every snapshot, which significantly reduces the I/O and long-term storage requirements. The dark matter distribution is, however, very useful in many applications, thus {\tt L-Gadget3} is able to store a subset of particles sampling homogeneously the initial Lagrangian distribution. 

\subsubsection{SubFind and Group finders}
\label{sec:subfind}


The identification of bound structures is a key aspect of $N$-body simulations, thus {\tt L-Gadget3} features a version of the {\tt SUBFIND} algorithm with several improvements. 

The first improvement is the ability to track subhalos on-the-fly across snapshots -- defining progenitor and descendants --, computing various additional quantities such as peak halo mass, peak maximum circular velocity, infall subhalo mass, and mass accretion rate, among others. These properties become useful when modelling galaxy formation within gravity-only simulations \citep[e.g.][]{ChavesMontero:2016,Moster:2018}.

The second improvement of our updated version of {\tt SUBFIND} is the use of the subhalo catalogue in the previous snapshot to better identify structures. In the original algorithm, particles are first sorted according to the local density, then when a saddle point is detected, the most massive group at that point is considered as the primary structure. This, however, can cause inconsistencies across time, as small changes can lead to fluctuations in the structure considered as primary. In our version of {\tt SUBFIND} instead, when a saddle point is detected, we consider as primary the substructure whose main progenitor is the most massive. This has proven to return more stable merger trees and quantities that are not local in time (e.g. peak maximum circular velocity). 

Finally, during their evolution, substructures can disappear from a simulation due to artifacts in the structure finder, finite numerical resolution, or because its mass falls below the resolution limit owing to tidal stripping. The last of our improvements to {\tt SUBFIND} is the ability to track all subhalos with no recognizable descendant, keeping the position and velocity of their most bound particle. This is a indispensable feature to correctly model the small-scale galaxy clustering in dark matter simulations \citep{GuoWhite:2014}.

At every output time, we store FoF groups and {\tt SUBFIND} subhalos with more than 10 particles. In total, there are approximately $129$ billion groups and $214$ billion subhalos in our outputs. This means our simulations are able to resolve halos with mass $5\times10^{10}\Msun$, and subhalos with a number density of $0.1\,h^{3}{\rm Mpc}^{-3}$ at $z=0$.

\subsection{Validation}

\subsubsection{Time and force resolution}
\label{sec:resolution}

\begin{figure}
\includegraphics[width=0.475\textwidth]{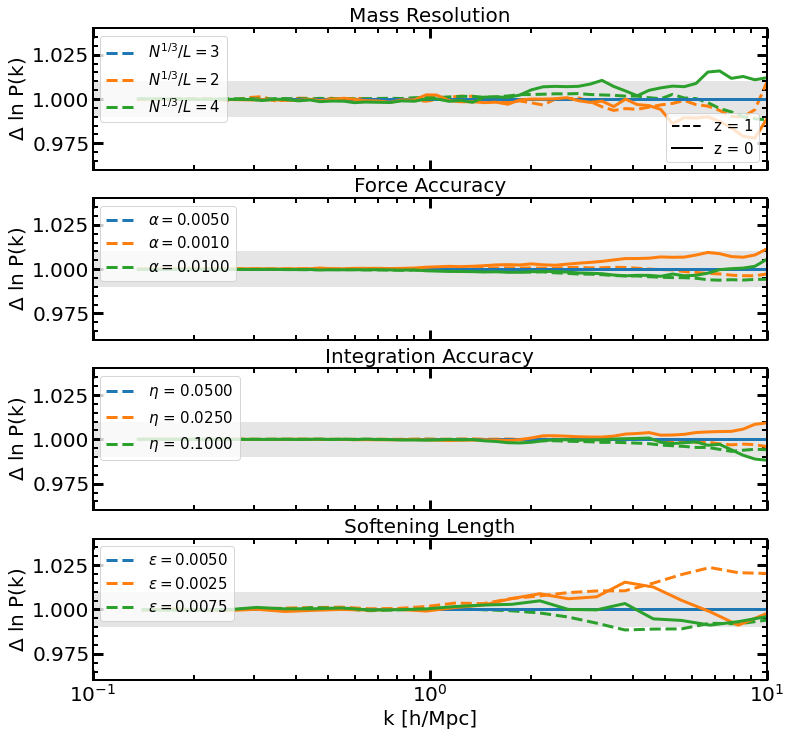}
\caption{The impact of numerical parameters in our simulated nonlinear mass power spectra at $z=0$ and $z=1$. We display fractional differences with respect to the measurements in a simulation that adopts the same accuracy parameters as our main BACCO simulations. The grey bands indicate a region of $\pm 1\%$. 
 \label{fig:pk_error}}
\end{figure}

In order to quantify the accuracy of our results, we have carried out a suite of small, $L=64\,\hMpc$ box, simulations where we systematically vary the numerical parameters around those adopted in the BACCO simulations. Specifically, we consider the time-integration accuracy, force calculation accuracy, softening length, and mass resolution. {\bf We adopt the parameters of the  {\tt nenya} simulation as the reference point for these tests.}

In Fig.~\ref{fig:pk_error} we compare the nonlinear power spectrum between these test runs and one that adopts the same numerical parameters and mean interparticle separation as our main BACCO simulations. Solid and dashed lines denote the results at $z=0$ and $z=1$, respectively. In all panels the grey shaded region indicates $\pm 1\%$ agreement.

In the top panel we display simulations with different mass resolutions, $N^{1/3}/L = [2, 3, 4]$. The main effect of mass resolution is how well small non-linear structures are resolved. We see no significant effect up to $k\sim5\ihMpc$ and a mild increase, of about 1\% in the power at $k\sim10\ihMpc$ when improving the mass resolution. 

The main source of inaccuracies in the force calculation are the terms neglected in the oct-tree multipole expansion. Specifically, {\tt Gadget} considers only the monopole contribution down to tree-nodes of mass $M$ and size $\ell$ that fulfill $\frac{GM^2}{r^2} \left( \frac{\ell}{r}\right) > \alpha |\vec{a}|$ for a particle at a distance $r$ and acceleration $|\vec{a}|$. Thus, the accuracy in the force calculation is controlled by the parameter $\alpha$. In the second panel we vary this parameter and see that the power spectrum is converged at a sub-percent level.

As in previous versions of {\tt Gadget}, time-steps in {\tt L-Gadget3} are computed individually for each particle as $\sqrt{2\eta\epsilon/|\vec{a}|}$, where $\epsilon$ is the softening and $|\vec{a}|$ is the magnitude of the acceleration in the previous timestep. The parameter $\eta$ therefore controls how accurately orbits are integrated. In the third panel we vary this parameter increasing/decreasing it by a factor of 4/2 with respect to our fiducial value, $\eta=0.05$. We see that the power spectrum varies almost negligibly with less than a $1\%$ impact up to $k\sim10\ihMpc$.  

Perhaps the most important degree of freedom in a numerical simulation is the softening length, $\epsilon$, a parameter that smooths two-body gravitational interactions (note that in {\tt Gadget}, forces become Newtonian at a distance $2.7\epsilon$). In the fourth panel we compare three simulations with 50\% higher and lower values of $\epsilon$ to values equal to $1/140$ and $1/35-th$ of the mean interparticle separation. On small scales we see that the amount of power increases systematically the lower the value of the softening length. In particular, our fiducial configuration underestimates the power by $1(2.5)$\% at $k\sim 5(10)\,\ihMpc$.

In summary, our results appear converged to better than 1\% up to $k \sim 5\ihMpc$, and to $\sim 3\%$ up to $k\sim 10\,\ihMpc$. The main numerical parameter preventing better convergence is the softening length. To mitigate its impact, we have adopted the following empirical correction to our power spectrum results

\begin{equation}
\label{eq:pk_correction}
P(k) \rightarrow P(k)\,\left[1 + 1.62\times10^{-2} ({\rm erfc}(\ln(k/\epsilon) - 3.1)-2)\right]
\end{equation}

\noindent where $\epsilon$ is the softening length in units of $\ihMpc$, and ${\rm erfc}$ is the complementary error function. We found this expression by fitting the effect seen in the bottom panel of Fig.~\ref{fig:pk_error}. We will apply this correction by default which brings the expected nominal accuracy of our power spectrum predictions to $\sim 1\%$ on all scales considered.

\subsubsection{The EUCLID comparison project}
\label{sec:euclid}

\begin{figure}
\includegraphics[width=0.45\textwidth]{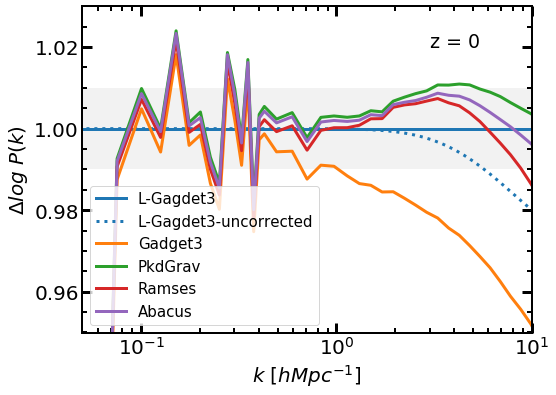}
    \caption{The nonlinear mass power spectrum at $z=0$ of the ``Euclid code comparison project''. Each coloured curve displays the predictions of a different $N$-body code, as indicated by the legend, and we display ratios relative to our simulation result corrected by finite numerical resolution. Note that all $N$-body codes agree to better than $1$\% precision up to $k \sim 5 \ihMpc$, with the exception of the original {\tt Gadget3} run presented in Schneider et al (2016).    
\label{fig:euclid_pk}}
\end{figure}

\begin{figure*}
\includegraphics[width=\textwidth]{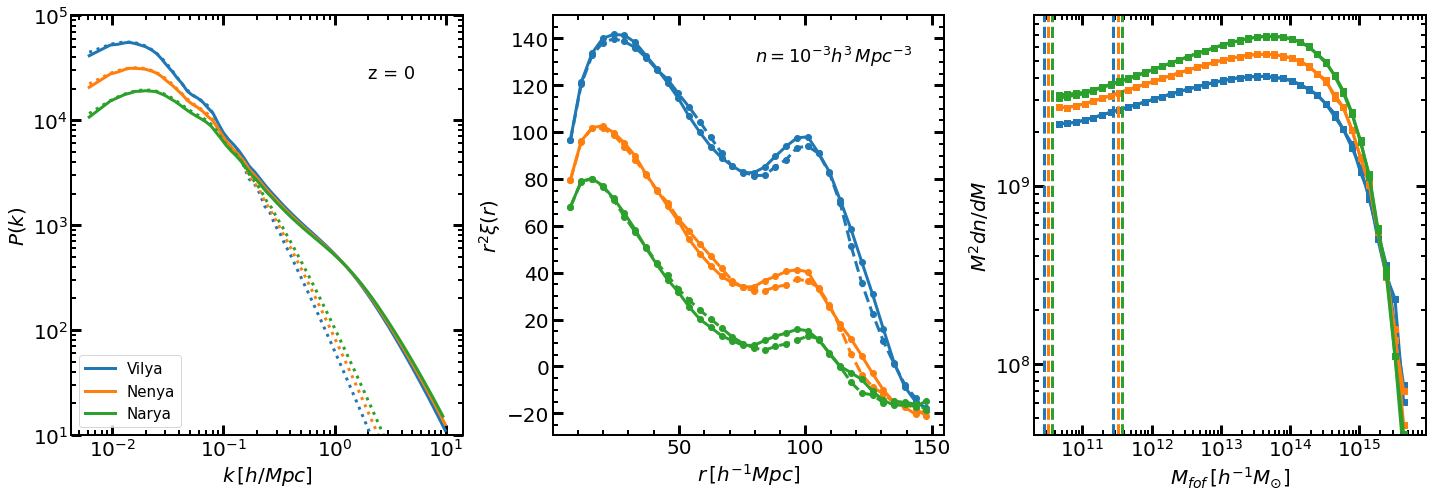}
\caption{ Predictions from the BACCO simulations, \vilya, \nenya, and \narya, at $z=0$. Left panel shows the nonlinear mass power spectrum as solid lines, and linear perturbation theory as dotted lines. The middle panel shows the redshift-space correlation function for subhalos with a number density of $10^{-3}\,h^{3}\,{\rm Mpc}^{-3}$, where solid and dashed lines show the results for each of the two opposite-phase simulations. The right panel shows the Friends-of-Friends halo mass function, with vertical dashed lines indicating the mass limit resolved with 10 and 100 particles.   
\label{fig:firstlook}}
\end{figure*}

To further validate our $N$-body code and quantify the accuracy of our numerical simulations, we have carried out the main simulation of the ``Euclid code comparison project'', presented in \cite{Schneider2016}. This simulation consists of $2048^3$ particles of mass $1.2\times10^9\,\Msun$ in a $500\,\hMpc$ box, and has been carried out with several $N$-body codes: {\tt RAMSES} \citep{Teyssier:2002}, {\tt PkdGrav3} \citep{Potter:2017}, {\tt Gagdet3} , and recently with {\tt ABACUS} by \cite{Garrison:2019}. 

Our realisation of this simulation adopts the same force and time-integration accuracy parameters as of main BACCO simulations, and the same softening length as \nenya, $\epsilon=5\,\hkpc$. The full calculation required $6230$ timesteps and took $3\times10^5$ CPU hours employing $1024$ MPI Tasks.

In Fig.~\ref{fig:euclid_pk} we compare the resulting power spectra at $z=0$. We display the ratio with respect to our {\tt L-Gadget3} results including the correction provided in Eq. \ref{eq:pk_correction}. For comparison, we display the uncorrected measurement as a blue dashed line. Note the spikes on large scales are caused by noise due to a slightly different $k$-binning in the spectra.

Our results, {\tt PkdGrav}, {\tt RAMSES}, and that of {\tt ABACUS} agree to a remarkable level -- they differ by less than 1\% up to $k\sim5\ihMpc$. This is an important verification of the absolute accuracy of our results, but it is also an important cross-validation of all these 4 $N$-body codes. Interestingly, the {\tt Gadget3} result presented in \cite{Schneider2016} is clearly in tension with the other $4$ codes. Since the underlying core algorithms in our code and in {\tt Gadget3} are the same, the difference is likely a consequence of numerical parameters adopted by \cite{Schneider2016} not being as accurate as those for the other runs. In the future, it will be important to conduct code comparison projects were numerical parameters are chosen so that each code provides results converged to a given degree.

\subsubsection{The BACCO simulations}

Having validated our numerical setup, we now present an overview of the results of our BACCO simulations at $z=0$ in Fig.~\ref{fig:firstlook}. 

In the left panel we show the nonlinear power spectrum in real space. Firstly, we see the low level of random noise in our predictions owing to the \paf\, method. On large scales, we have checked that our results agree at the $0.5\%$ level with respect to the linear theory solution, which we compute for each cosmology using the Boltzmann code {\tt CLASS} \citep{Lesgourgues2011}. Also on large scales, we see that the three simulations display significantly different power spectra, despite the three of them having identical values for $\sigma_8$. This is mostly a consequence of their different primordial spectral index, $n_s$. In contrast, on scales smaller than $k \sim 0.2\ihMpc$, the spectra become much more similar as their linear spectra also do (shown by dotted lines).

In the middle panel we show the monopole of the redshift-space correlation functions for subhalos with a spatial number density equal to $10^{-3}\,h^3{\rm Mpc}^{-3}$ selected according to their peak maximum circular velocity. This is roughly analogous to a stellar mass selection above $5\times10^{10}\Msun$. We see a significant difference in the correlation amplitude among simulations. This hints at the potential constraining power of LSS if a predictive model for the galaxy bias is available. We display each of the two \paf\ simulations in each cosmology as solid and dashed lines. Unlike in the power spectrum plot, the effect of pairing the initial phase fields is visible.

Finally, in the right panel of Fig.~\ref{fig:firstlook} we display the mass function of  halos identified by the FoF algorithm. 
As in previous cases, there are clear differences among cosmologies. The minimum halo mass in our simulations is $\sim 3-4\times10^{10}\,\Msun$, which should suffice to model galaxies with star formation rates above $10\,\Msun$/year at $z\sim1$ \citep{Orsi:2018}, as expected to be observed by surveys like EUCLID, DESI, or J-PAS \citep{Favole:2017}.

In the next section we will employ our BACCO simulations to predict the nonlinear mass power spectrum as a function of cosmology.

\section{Nonlinear mass power spectrum Emulator} \label{sec:emulator}

Our aim is to make fast predictions for the nonlinear power spectrum over the whole region of currently-viable cosmologies. In this section we describe how we achieve this by  building (\S\ref{sec:emu_build}) and testing of a matter power spectrum emulator (\S\ref{sec:emu_test}).

Our basic strategy is the following:
\begin{itemize}
\item[•] First, we define a target region in cosmological parameter space (\S\ref{sec:emu_params}) and iteratively select a set of training points that minimise the emulator uncertainty (\S\ref{sec:emu_training}). 
\item[•] Second, we use the cosmology-rescaling approach over the outputs of our BACCO simulations (\S\ref{sec:emu_scaling}) to predict the power spectra in those training cosmologies (\S\ref{sec:emu_pk}). 
\item[•] Finally, employ those measurements to build an emulator for the power spectra based on either Gaussian Processes or Neural Networks (\S\ref{sec:emu_gaussian}). 
\end{itemize}

We test the performance and accuracy of our emulator (\S\ref{sec:emu_accuracy}) and compare it against widely-used methods to predict the nonlinear power spectrum (\S\ref{sec:emu_comparison}). 

\subsection{Building the emulator}
\label{sec:emu_build}

\begin{figure*}
\includegraphics[width=0.95\textwidth]{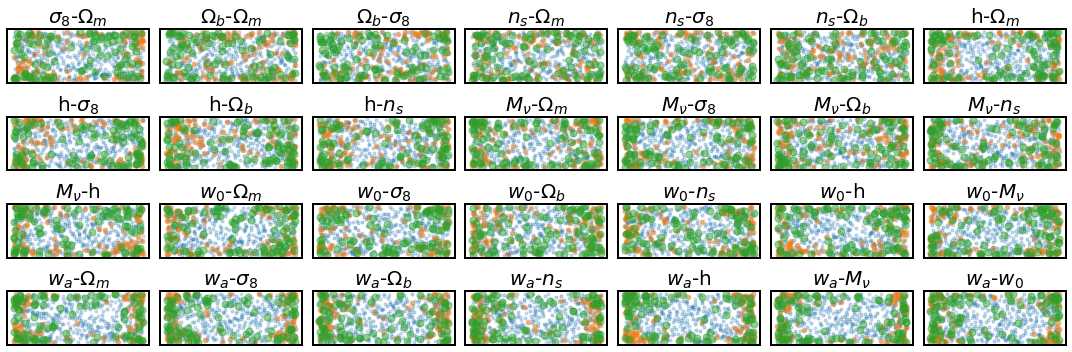}
\caption{Distribution of cosmologies employed to build our BACCO emulator. Blue symbols display the location of cosmologies in our initial training set, whereas orange and green symbols display those subsequently selected by our iterative method. In each plot, the limits coincide with the full range of values considered (c.f. Eq~\ref{eq:par_range}).
\label{fig:emu_coords} }
\end{figure*}

\subsubsection{The parameter space}
\label{sec:emu_params}

In this paper we aim to cover 8 parameters of the $\Lambda$CDM model extended with massive neutrinos and dynamical dark energy. Specifically, we consider the parameter range:

\begin{eqnarray}
\label{eq:par_range}
\sigma_8                  &\in& [0.73, 0.9] \nonumber\\
\Omega_{\rm m}            &\in& [0.23, 0.4] \nonumber\\
\Omega_b                  &\in& [0.04, 0.06] \nonumber\\
n_s                       &\in& [0.92, 1.01]\\
h\,[100\,{\rm km}\,{\rm s^{-1}} {\rm Mpc^{-1}}]  &\in& [0.6, 0.8] \nonumber\\
M_{\nu}\,[{\rm eV}]       &\in& [0.0, 0.4] \nonumber\\
w_{0}                     &\in& [-1.15, -0.85] \nonumber\\
w_{a}                     &\in& [-0.3, 0.3] \nonumber
\end{eqnarray}


\noindent where $M_\nu$ is the total mass in neutrinos, $\sigma_8$ is the r.m.s. linear {\it cold mass} (dark matter plus baryons) variance in $8\hMpc$ spheres, and $w_0$ and $w_a$ define the time evolution of the dark energy equation of state: $w(z) = w_0 + (1 - a) w_a$. The parameter range we consider for ($\sigma_8$, $\Omega_m$, $\Omega_b$, $n_s$) corresponds to a $\sim10\sigma$ region around the best-fit parameters of the analysis of \cite{Planck:2014}. For the dimensionless Hubble parameter, $h$, we expand the range to cover a $4\sigma$ region around current low-redshift measurements from supernovae data \citep{Riess:2019}.

We assume a flat geometry, i.e. $\Omega_k = 0$ and $\Omega_m + \Omega_w + \Omega_{\nu} = 1$. We keep fixed the effective number of relativistic species to $N_{\rm eff}=3.046$, and the temperature of the CMB $T_{\rm CMB} = 2.7255$ K and neglect the impact of radiation (i.e. $\Omega_r = 0$). Note, however, that it is relatively straightforward to relax these assumptions and include in our framework varying curvature, relativistic degrees of freedom, or other ingredients.

For comparison, we notice that the range of parameters covered by our emulator is approximately twice as large as that of the {\tt Euclid emulator} (with the exception of $w_0$, which is similar), which implies a parameter space volume $\sim200$ times larger, and that it covers cosmological parameters beyond the minimal $\Lambda$CDM: dynamical dark energy and massive neutrinos. In contrast, our parameter space is similar to that considered by the {\tt Mira-Titan} Universe project \citep{Heitmann:2016,Lawrence:2017}, however, as we will discuss next, we cover the space with approximately 20 times the number of sampling points.

\subsubsection{Training Cosmology Set}
\label{sec:emu_training}

We now define the cosmologies with which we will train our emulator. This is usually done by sampling the desired space with a Latin-Hypercube \citep[e.g.][]{Heitmann:2006}. We adopt a slightly different strategy based on the idea of iterative emulation of \cite{Pellejero-Ibanez:2020} \citep[see also][]{Rogers2019}, where we preferentially select training points in regions of high emulator uncertainty. 

Let us first define the quantity we will emulate: 

\begin{equation}
Q(k, z) \equiv \log[ P(k,z)/P_{\rm linear}^{\rm smeared-BAO}(k,z) ],
\end{equation}

\noindent where $P(k)$ is the measured nonlinear power spectrum of {\it cold} matter (i.e. excluding neutrinos). $P_{\rm linear}^{\rm smeared-BAO}$ is the linear theory power spectrum where the BAO have been smeared out according to the expectations of perturbation theory. Specifically, we define:

\begin{equation}
P_{\rm linear}^{\rm smeared-BAO} \equiv P_{\rm linear}\,G(k) + P_{\rm linear}^{\rm no-BAO}\,[1 - G(k)]
\end{equation}

\noindent where $P_{\rm linear}^{\rm no-BAO}$ is a version of the linear theory power spectrum without any BAO signal. Operationally, this is obtained by performing a discrete sine transform, smooth the result, and return to Fourier space by an inverse transform \citep{Baumann:2018}. $G(k) \equiv \exp[-0.5\,k^2 / k_*]$, with $k_*^{-1} \equiv (3\pi^2)^{-1} \int {\rm d}k \,P_{\rm linear}$.

Emulating the quantity $Q(k)$, rather than the full nonlinear power spectra, reduces significantly the dynamical range of the emulation, simplifying the problem and thus delivering more accurate results \cite[see e.g.][]{Knabenhans:2019,Giblin:2019}. We emphasise, however, that these transformations are simply designed to improve the numerical stability of our predictions -- our results regarding large scales and BAO are still uniquely provided by our simulation results.   

To build our training set, we first construct a Latin-Hypercube with 400 points. We then build a Gaussian Process (GP) emulator for $Q$, which provides the expectation value and variance for the emulated quantity (c.f. \S\ref{sec:emu_gaussian} for more details). We then evaluate the GP emulator over a Latin-Hypercube built with 2,000 points, and select the 100 points expected to have the largest uncertainty in their predictions. We add these points to our training set and re-build the GP emulator. We repeat this procedure 4 times.

We display the final set of training cosmologies, comprised of 800 points, in  Fig.~\ref{fig:emu_coords}. Blue symbols indicate the initial training cosmologies, whereas orange and green symbols do so for the 2nd and 4th iteration, respectively.  We can appreciate that most points in the iterations are located near the boundaries of the cosmological space, which minimise extrapolation. 

It is worth mentioning that we sample our space significantly better than typical emulators. For instance, the {\tt Mira-Titan} and {\tt Euclid Emulator} projects employ $36$ and $100$ simulations, respectively. This implies that we can keep the emulator errors under any desired level, which could be problematic otherwise, as parameter-dependent uncertainties in models, could in principle bias parameter estimates. On the other hand, sampling points can be designed optimally so that emulation errors are much smaller than a naive Latin Hypercube \citep{Heitmann:2014,Rogers2019}, however, this can be only be done optimally for a single summary statistic: what could be optimal for the matter power spectrum, is not necessarily the best for, e.g, the quadrupole of the galaxy power spectrum.

\subsubsection{Cosmology rescaling}
\label{sec:emu_scaling}

Our next step is to predict nonlinear structure for each of our training cosmologies. For this, we employ the latest incarnation of the cosmology-rescaling algorithm, originally introduced by \cite{AnguloWhite:2010}.

In short, the cosmology-rescaling algorithm seeks a coordinate and time transformation such that the linear rms variance coincides in the original and target cosmologies. This transformation is motivated by extended Press-Schechter arguments, and returns highly accurate predictions for the mass function and properties of collapsed objects. On large scales, the algorithm uses 2nd-order Lagrangian Perturbation Theory to modify the amplitude of Fourier modes as consistent with the change of cosmology. On small scales, the internal structure of halos is modified using physically-motivated models for the concentration-mass-redshift relation \citep[e.g.][]{Ludlow:2016}. 

The accuracy of the cosmology algorithm has been tested by multiple authors \citep{Ruiz:2011,AnguloHilbert:2015,Mead:2014,Mead:2015,Renneby:2018,Contreras:2020}, and it has been recently extended to the case of massive neutrinos by \cite{Zennaro:2019}. Specifically, by comparing against a suite of $N$-body simulations, \cite{Contreras:2020} explicitly showed that the cosmology rescaling achieves an accuracy of $\lesssim 3\%$ up to $k = 5\ihMpc$ over the same range of cosmological parameter values we consider here (c.f. Eq.~\ref{eq:par_range}). The largest errors appear on small scales and for dynamical dark energy parameters but, when restricted to the 6 parameters of the minimal $\Lambda$CDM model, the rescaling returns $1\%$ accurate predictions. Note that the level of accuracy is set by the performance of current models for the concentration-mass-redshift relation (which usually are not calibrated for beyond $\Lambda$CDM parameters), and future progress along those lines should feedback into higher accuracy for our emulator. 

For our task at hand, we first split the parameter space into three disjoint regions where \nenya, \narya, or \vilya\ will be employed \citep{Contreras:2020}. Then, we load in memory a given snapshot of a given simulation and then rescale it to the corresponding subset of the 800 cosmologies. We employ 10 snapshots per simulation. 

The full rescaling algorithm takes approximately 2 minutes (on 12 threads using OpenMP parallelization) per cosmology and redshift. Thus, all the required computations for building the emulator required approximately $10,000$ CPU hours -- a negligible amount compared to that employed to run a single state-of-the-art $N$-body simulation.

\subsubsection{Power Spectrum Measurements}
\label{sec:emu_pk}

We estimate the power spectrum in our rescaled simulation outputs using Fast Fourier Transforms. We employ a cloud-in-cells assignment scheme over two interlaced grids \citep{Sefusatti:2016} of $N=1024^3$ points. To achieve efficient measurements at higher wavenumbers, we repeat this procedure after folding the simulation box $4$ times in each dimension. We measure the power spectrum in $160$ evenly-spaced bins in $\log{k}$, from $k=0.01$ to $5\ihMpc$, and transition from the full to the folded measurement at half the Nyquist frequency. We have explicitly checked that this procedure returns sub-percent accurate power spectrum measurements over the full range of scales considered. 

Although our BACCO simulations have high mass resolution, for computational efficiency, hereafter we will consider a subset of $1080^3$ particles (uniformly selected in Lagrangian space) as our dark matter catalogue. This catalogue, however, is affected by discreteness noise. For a Poisson sampling of $N$ points over a box of side-length $L$, the power spectrum receives a contribution equal to $(L/N)^3$. However, this might not be an accurate estimate for our discreteness noise since our sampling is homogeneous in Lagrangian coordinates. We have estimated its actual contribution as a third-order polynomial by comparing the spectra of the full and diluted samples at different redshifts. We found that the amplitude is proportional to $\sim0.6$ times the Poisson noise at $z=0$, and that it progressively decreases in amplitude at higher redshifts, to reach $\sim0.2$ times the Poisson noise at $z=2$. 

All our power spectrum measurements will be corrected by discreteness noise by subtracting the term described above, and further corrected for finite numerical accuracy following section \S\ref{sec:resolution}. However, this procedure is not perfect, and we still detect $\sim2$\% residuals at $z\sim1$ at $k\sim5\ihMpc$. This will contribute to uncertainties in our emulator, which, however, are smaller than systematic uncertainties induced by the cosmology scaling.

\subsubsection{Emulator Data}

\begin{figure}
\includegraphics[width=0.45\textwidth]{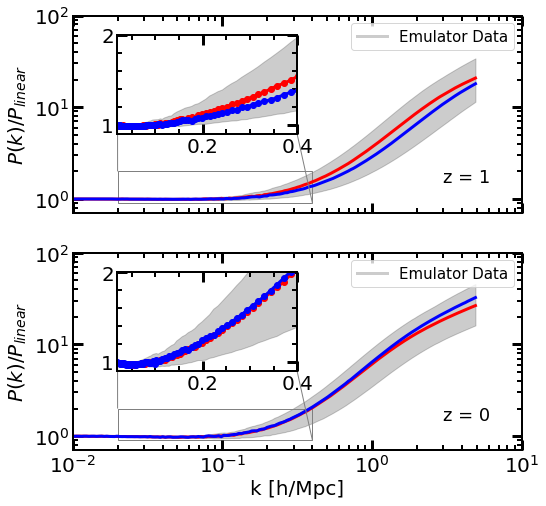}
\caption{Nonlinear power spectra at $z=0$ and $z=1$ over the linear theory expectations, as predicted by rescaled BACCO simulations. Grey regions indicate the range covered by the training set employed to construct our emulator. Blue and red lines display two particular, randomly-chosen, measurements. 
\label{fig:emu_data}}
\end{figure}

In total, we employ more than 16,000 power spectrum measurements; $400+4\times100$ training cosmologies at approximately 10 different cosmic times for two paired simulations. Grey shaded regions in Fig.~\ref{fig:emu_data} show the range covered by this data at $z=0$ and $z=1$, with the inset focusing on the range of wavemodes where baryonic acoustic oscillations are found. Blue and red lines show two randomly-chosen cosmologies.

On the largest scales, we can see that our rescaled simulations agree almost perfectly with linear theory. Although expected, it provides further validation of the dataset. On intermediate scales, we can see a lack of any oscillatory behaviour -- better appreciated in the figure insets. This is a consequence of our $P_{\rm linear}^{\rm smeared-BAO}$ correctly capturing the nonlinear smearing of baryonic acoustic oscillations caused mostly by large-scale flows. This is true at both $z=0$ and $z=1$, which confirms the correct redshift dependence of $P_{\rm linear}^{\rm smeared-BAO}$. On small and nonlinear scales, we see an increase of more than one order of magnitude with different cosmologies differing by even factors of 3-4. 

\subsubsection{Principal Component Analysis}

\begin{figure}
\includegraphics[width=0.465\textwidth]{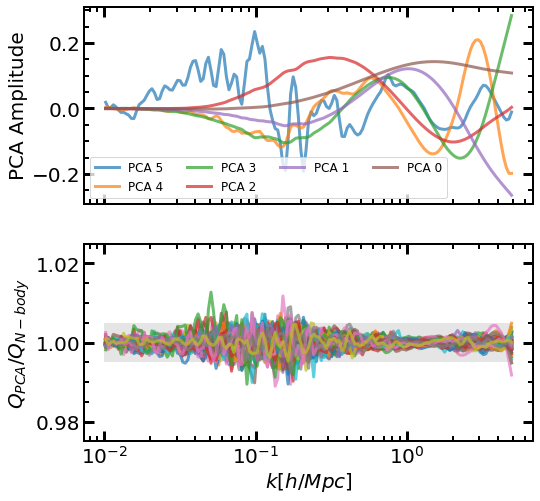}
\caption{Top: The amplitude of the first $6$ vectors of a principal component analysis of our power spectrum data. Bottom: Ratio between the original and PCA-reconstructed $Q \equiv \log( P/P_{\rm linear}^{\rm smeared-BAO})$ for a random $2\%$ of our data at $z=0$. The grey band indicates a region of $\pm 0.5\%$, which can be considered as an indication of the statistical noise in our data. 
\label{fig:emu_pca}}
\end{figure}

\begin{figure*}
\includegraphics[width=0.95\textwidth]{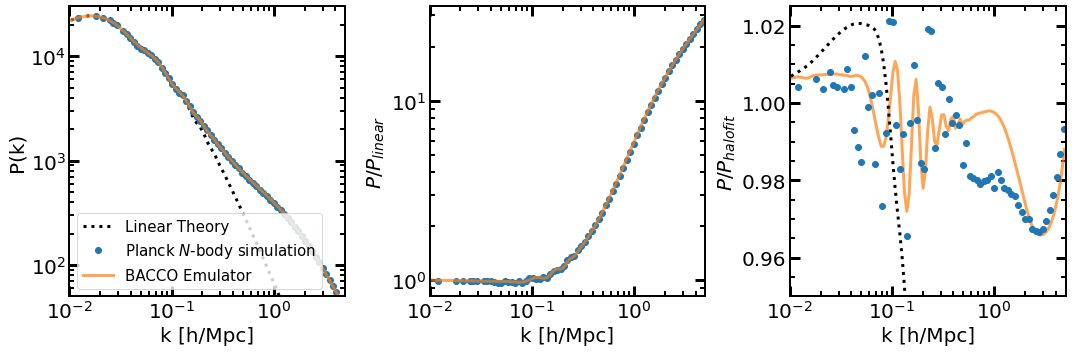}
\caption{Comparison between the nonlinear power spectrum at $z=0$ as predicted by a full $N$-body simulation and our BACCO emulator for a set of cosmological parameters consistent with recent analyses of the Planck satellite data. The left panel shows the full power spectrum, whereas middle and rightmost panel show the same data relative to the expectations of linear theory or halofit, respectively. The emulator and simulation predictions agree to better than $\sim2$\% over the full range of scales considered. 
\label{fig:emu_example}}
\end{figure*}

To reduce the dimensionality of our power spectrum measurements, we have performed a principal components (PC) analysis over our whole dataset, after subtracting the mean. We have kept the 6 $k$-vectors with the highest eigenvalues, which together can explain all but $10^{-3}$ of the data variance. 

In the top panel of Fig.~\ref{fig:emu_pca} we show the amplitude of these PCs, as indicated by the legend. The most important vector is a smooth function of wavenumber and roughly captures the overall impact of nonlinear evolution. Subsequent vectors describe further modifications, but none of them shows significant oscillations, which indicates that $P_{\rm linear}^{\rm smeared-BAO}$ accurately models the BAO as a function of cosmology and redshift. It is worth noting that only the 6th vector displays any noticeable noise, owing to the highly precise input dataset.

In the bottom panel of the Fig.~\ref{fig:emu_pca} we display the ratio of the full $Q\equiv \log(P/P_{\rm linear})$ over that reconstructed using the first PCs aforementioned. We show the results for a random 2\% of the power spectrum in our training data. We can see that almost all of them are recovered to better than $0.5\%$. It is interesting to note that the residuals, although increase on intermediate scales, are mostly devoid of structure, which suggests that including additional PCs will simply recover more accurately the intrinsic noise in our dataset rather than systematic dependencies on cosmology. 

To confirm this idea, we have repeated our emulation but keeping twice as many PCs. Although the description of the values of Q in our dataset gained accuracy (decreasing the differences down to $\sim 0.25\%$), the performance of the emulator at predicting other cosmologies did not increase. This supports the idea that PCs beyond the 6-{\it th} are simply capturing particular statistical fluctuations in the training set, rather than cosmology-induced features. It is interesting to note that this also means that all nonlinear information of the power spectrum beyond the BAO can be described by only $6$ numbers.

\subsubsection{Emulation}
\label{sec:emu_gaussian}

\begin{figure}
\includegraphics[width=0.45\textwidth]{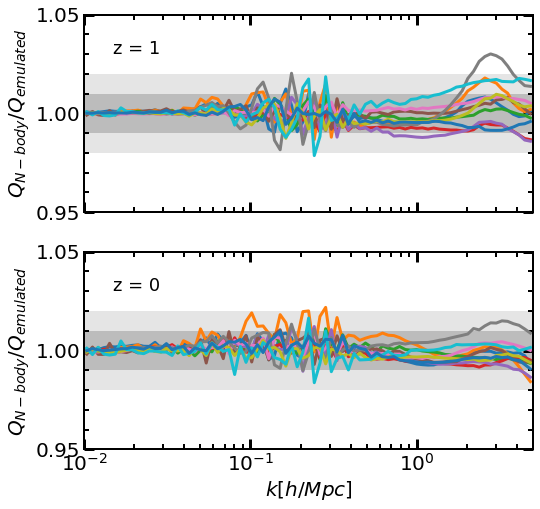}
\caption{Comparison between $Q \equiv \log(P/P_{\rm linear})$ predicted by our emulator and that computed directly by scaling our BACCO simulations to the desired cosmology. Top and bottom panels display results at $z=1$ and $z=0$, respectively, for 10 cosmologies that span our whole parameter space.   
\label{fig:emu_test_pk}}
\end{figure}

To interpolate between our training data, and thus predict $Q(k,z)$ for any cosmology, we will construct two emulators: one built by employing Gaussian Processes Regression, and another one built by training a Neural Network. 

\subsubsection*{Gaussian Processes}

In short, Gaussian processes assume that every subset of points in a given space is jointly Gaussian distributed with zero mean and covariance $K$. The covariance is {\it a priori} unknown but it can be estimated from a set of observations (e.g. our training set). Once the covariance is specified, the Gaussian process can predict the full probability distribution function anywhere in the parameter space.

In our case, we measure the amplitude associated to each PC in each training set. We then build a separate Gaussian Process for each of our PCs, using the package {\tt GPy} \citep{gpy2014}. We assume the covariance kernel to be described by a squared exponential, with correlation length and variance set to the values that best describe the correlation among our data, found by maximising the marginal likelihood.

\subsubsection*{Neural Network}

For regression tasks, neural networks perform remarkably well, serving as a highly-precise approximation to almost any continuous function. Thus, they provide an alternative for emulating $Q(k,z)$ \citep[see e.g.][]{Agarwal:2014,Kobayashi:2020}.

In our case, we found that a feed-forward neural network with a simple architecture allowed us to obtain an accurate emulation. Specifically, we employed a fully-connected network with 2 hidden layers, each with 2000 neurons. For the activation function -- which takes our input vectors and transforms them non-linearly -- we use of Rectified Linear Units (ReLUs) which are commonly used in machine learning. The training of the network is done by using the adaptive stochastic optimization algorithm {\tt Adam} \citep{kingma2014adam}, with a default learning rate of $10^{-3}$. For the implementation of the neural network we made use of the open-source neural-network python library, {\tt Keras} \citep{chollet2015keras}. 

We employ the PCA-reconstructed $Q(k,z)$ as our training data for the neural network (NN), so we can compare its results to those of the Gaussian process in equal terms. We randomly selected 5\% of our power spectra as our validation set, and employ the remaining 95\% as our training data. We define the loss function as the mean squared error function (MSE), and employed 2000 epochs for the training, after which our results appeared converged. At this point, the value of the loss function evaluated in the training and validation sets differed by less than $\sim50\%$, which suggest no overfitting. We note that we also tried the so-called batch-normalization and data-dropouts, without finding significant improvements in our predictions.

For a given target cosmology in our parameter space, we can predict the amplitude associated to each PC and then reconstruct the full $Q(k,z)$ vector.  We found both emulators displaying similar performances: we compared their predictions at 1000 random locations within our parameter space, and differences were typically smaller than $1-2$ percent. Both emulators display similar computational performances, about $200$ milliseconds for predicting $Q$ on any cosmology.

\begin{figure*}
\includegraphics[width=0.95\textwidth]{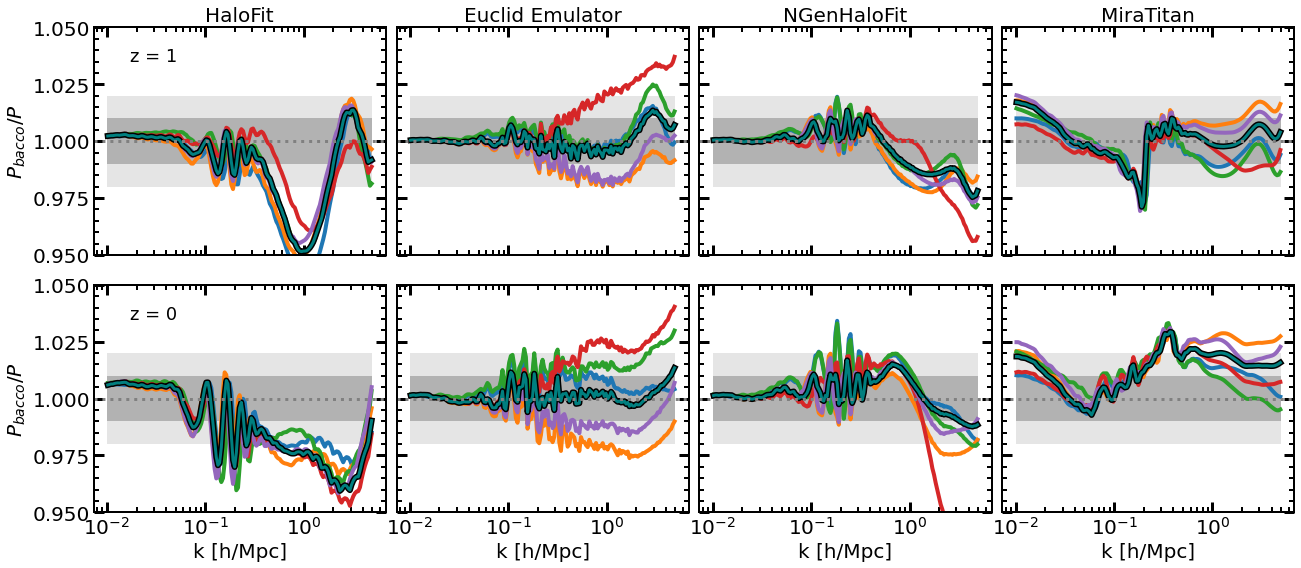}
\caption{Comparison between nonlinear power spectrum predicted by our emulator and by four other methods. From left to right, panels display a comparison against {\tt Halofit}, {\tt Euclid Emulator}, {\tt NGenHaloFit}, and {\tt MiraTitan}. Top and bottom rows show results for $z=1$ and $z=0$, respectively. Each coloured curve represents a different cosmology within the parameter space where the {\tt Euclid Emulator} has been calibrated. We highlight the ``Euclid reference'' cosmology as a thick line. Grey bands indicate $\pm 1$ and $\pm 2$\% regions.
\label{fig:emu_comparison}}
\end{figure*}

\begin{figure}
\includegraphics[width=0.45\textwidth]{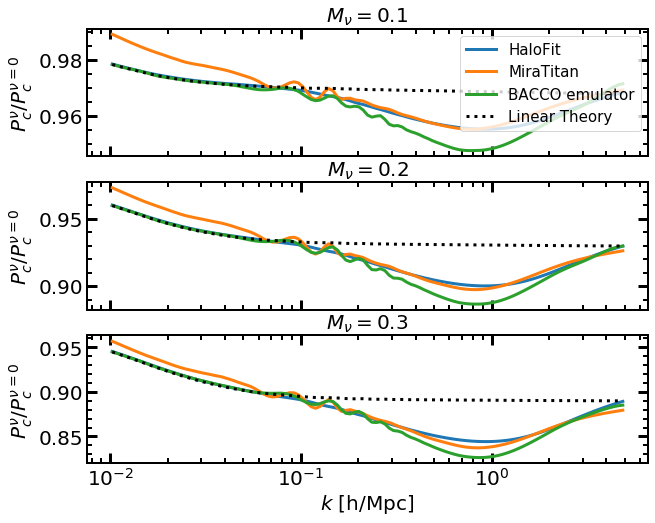}
\caption{ The ratio between the {\it cold} matter power spectra in massive neutrino cosmologies, $P^{\nu}_c$, and the case without any massive neutrino, $P^{\nu = 0}_c$. We show the predictions of linear theory (dotted line) and of three models: {\tt HaloFit}, {\tt Mira-Titan} emulator and our BACCO emulator.     
\label{fig:emu_nu}}
\end{figure}

\subsection{Testing the emulator}
\label{sec:emu_test}

\subsubsection{A first example}

We present a first look at our emulator results in Fig.~\ref{fig:emu_example}, where we show the nonlinear power spectrum at $z=0$ for a cosmology consistent with a recent analysis of Planck data. Specifically, we consider: $\sigma_8=0.8102$, $\Omega_m=0.30964$, $\Omega_b=0.04897$, $n_s=0.9665$, $h=0.6766$, $M_{\nu}=0.0$, $w_0=-1.0$, $w_a=0.0$. The left panel shows the full power spectrum, the middle panel do so relative to linear theory, and the rightmost panel relative to {\tt HaloFit} calibrated by \cite{Takahashi:2012}.

For comparison, we include as orange circles, the results of a full $N$-body simulation with the same mass resolution and numerical parameters as those of our main BACCO suite, but in a smaller box, $L=512\,\hMpc$. Also as in our BACCO suite, the initial conditions have been \paf\ for this simulation.

Overall, we can see that the BACCO emulator and the $N$-body simulation agree to a remarkable level, being indistinguishable by eye in the leftmost and middle panels. In particular, on large scales, both agree with linear theory (which only can be appreciated thanks to the \paf\ initial conditions); on intermediate scales, both also predict a BAO featured smeared out compared to linear theory.

In the rightmost panel we can see these aspects in more detail. Firstly, we note that the simulation and emulator results agree to about 1\% on all the scales considered. This is consistent with the expected accuracy of the cosmology rescaling method, but also note that due to its somewhat small volume, the cosmic variance in the $N$-body simulation results are not negligible. It is also interesting to note the systematic disagreement with the predictions from {\tt halofit}.  In following subsections we will explore these differences further.

\subsubsection{Accuracy}
\label{sec:emu_accuracy}

To start testing the accuracy of our emulator, we have defined 10 cosmologies distributed over the target parameter space (c.f.\S\ref{sec:emu_params}) using a Latin hypercube. We then rescale our BACCO outputs to those parameters and compare the results against our emulation predictions. This essentially tests how accurate the PCA decomposition and emulation via Gaussian Process Regression are.

In Fig.~\ref{fig:emu_test_pk} we show the ratio of the emulated to the rescaled nonlinear power spectrum at $z=0$ and $z=1$ for the 10 cosmologies mentioned before (we recall that neither $z=0$ nor $z=1$ were explicitly included in the emulator). We can see that on all scales our results are better than $\pm2\%$ percent, and that most of the cosmologies are predicted to better that 1\% at $z=0$. Although this is already a high accuracy, we note that more and more rescaled results can be added over time to progressively improve the quality of the emulation. We recall that this level of uncertainty is comparable to the accuracy expected by the cosmology rescaling algorithm: $1\%$ for parameters of the minimal $\Lambda$CDM and $\sim3\%$ when considering massive neutrinos and dynamical dark energy.


\subsubsection{Comparison with HaloFit, EuclidEmulator, and NgenHaloFit}
\label{sec:emu_comparison}

We now compare our emulation results against four widely-used methods to predict the nonlinear evolution: {\tt HaloFit} \citep{Takahashi:2012}, the {\tt Euclid Emulator} \citep{Knabenhans:2019}, {\tt NGenHalofit} \citep{SmithAngulo:2019}, and {\tt Mira-Titan} \citep{Lawrence:2017}. Since not all of them have been calibrated over the whole parameter space covered by our emulator, we have restricted the comparison to the volume covered by the {\tt Euclid Emulator}. We note that of our 800 training cosmologies, only 2 of them fall within this parameter volume.

In Fig.~\ref{fig:emu_comparison} we display the ratio of those predictions to that of our BACCO emulator. Coloured lines show 10 cosmologies set by a latin hypercube inside the {\tt Euclid Emulator} parameter space. In addition, we show as a heavy line the Euclid reference cosmology: $\Omega_{\rm cdm} = 0.26067$, $\sigma_8 = 0.8102$, $\Omega_{\rm b} = 0.04897$, $n_s = 0.9665$, $h = 0.6766$, $M_{\nu} = 0$, $w_0=-1$, $w_a=0$, employed by \citep{Knabenhans:2019}.

On large scales, $k < 0.08\ihMpc$, our emulator agrees almost perfectly with {\tt NGenHalofit} and the {\tt Euclid Emulator}. {\tt Halofit}, on the other hand shows a small constant power deficit, whereas the {\tt Mira-Titan} displays a weakly scale-dependent offset. Over this range of scales, {\tt Mira-Titan} is given by TimeRG perturbation theory   \citep{Pietroni:2008,Upadhye:2014}, which might indicate inaccuracies in that approach.

On intermediate scales, $0.08 < k/[\ihMpc] < 0.5$, our results agree very well with those of {\tt NGenHalofit}; for almost all cosmologies the differences are within the expected accuracy of our emulator. In contrast, we see that {\tt Halofit} overestimates the amount of power by about 2\% at $z=0$ and it does not correctly captures the BAO nonlinear smearing \citep[as was already noted by][]{Knabenhans:2019}. On the other hand, {\tt MiraTitan} underestimates the power by up to 2.5\% at $z=0$, and displays a strong feature at $k \sim 0.2\ihMpc$ at $z=1$. Over this range of scales {\tt MiraTitan} employs a suite of low-resolution, large volume simulations, thus the disagreement could be due to an insufficient numerical accuracy in their simulations. 

On small scales, $k > 0.5\ihMpc$, we continue to see differences among the three methods. For halofit, {\tt NGenHaloFit} and {\tt Mira-Titan}, they are systematic, roughly independent of cosmology, and decrease at higher redshift. A possibility is  that this could be originated by the different numerical accuracy of the underlying simulations. Specifically, the simulations used by \cite{SmithAngulo:2019} employ a softening length $\epsilon=8\hkpc$, which according to Eq.~\ref{eq:pk_correction} is expected to produce an underestimation of 1.5\% at $k=5\ihMpc$. In addition, the transition between their high and low-resolution runs occurs at $k\sim0.6\ihMpc$ which might be related to the deficit we observe at $k \sim 0.7-0.7\ihMpc$ at $z=0$.

In contrast, the differences with respect to {\tt Euclid Emulator} vary significantly for different cosmologies. Specifically, for the Euclid Reference Cosmology, the agreement is sub-percent on all scales. However, for our other test cosmologies, differences have a spread of $\sim5$\%, even at $k < 1\ihMpc$. We do not see this behavior with other methods, which might suggest that there are significant uncertainties in the emulated power spectra of the {\tt Euclid Emulator} beyond their quoted precision.

Finally, we note that there is a 1\% ``bump'' at $k\sim3\ihMpc$ in all $z=1$ panels, which is originated by our imperfect shot-noise correction. 

The previous comparison was done in a rather restricted cosmological parameter volume, which served as a strong test of our accuracy. However, one of the biggest advantages of our method is the ability to predict much more extreme cosmologies even with non-standard ingredients. We provide an example of this next.

In Fig. \ref{fig:emu_nu} we show the predictions for the effects of massive neutrinos on the $z=0$ cold matter (baryons plus dark matter) power spectrum. We display the ratio between cases with various neutrino masses, $P_{c}^{\nu}$, relative to that without massive neutrinos $P_{c}^{\nu}$. In all cases we use $A_s = 2.1\times10^{-9}$ and fix all the other cosmological parameters to those of the ``Euclid reference cosmology''. We only display the predictions of {\tt HaloFit} and {\tt Mira-Titan}, since {\tt NGenHaloFit} nor the {\tt Euclid Emulator} have been calibrated in the case of massive neutrinos.

We can see that on large scales, our predictions agree with both linear theory and {\tt HaloFit}. For very massive neutrino cases, there is also a good agreement with {\tt MiraTitan}, but there is a significant disagreement for $M_{\nu} = 0.1$, this might be caused by the scale-dependent features on large scales seen in the previous figure. On intermediate scales, all predictions also agree on the broadband shape of the neutrino-induced suppression, but our emulator is also able to capture the slightly different BAO suppression expected when massive neutrinos are present. On small scales, all three methods describe the well-known neutrino-induced spoon-like suppression, disagreeing slightly on the magnitude of the maximum suppression. Note however, we expect our emulator to predict more precisely the full shape of the nonlinear power spectrum, owing to the systematic uncertainties in {\tt HaloFit} and {\tt Mira-Titan} discussed before.

\section{Summary}
\label{sec:summary}

In this paper we have presented the BACCO simulation project: a framework that aims at delivering high-accuracy predictions for the distribution of dark matter, galaxies, and gas as a function of cosmological parameters.

The basic idea consists in combining recent developments in numerical cosmology -- $N$-body simulations, initial conditions with suppressed variance, and cosmology-rescaling methods -- to quickly predict the nonlinear distribution of matter in a cosmological volume. The main advantage of our approach is that it requires only a small number of $N$-body simulations, thus they can be of high resolution and volume. This in turn allows sophisticated modelling of the galaxy population (for instance in terms of subhalo abundance matching, semi-empirical or semi-analytic galaxy formation model), and of baryons (including the effects of cooling, star formation and feedback) in the mass distribution.

In this paper we have presented the main suite of 
{\bf gravity-only}
simulations of the BACCO project. These consist in 3 sets of \paf\ simulations, each of them of a size $L=1440\,\hMpc$ and with $80$ billion particles. Their cosmologies were carefully chosen so that they maximise the accuracy of our predictions (Fig.~\ref{fig:density_field} and Table~\ref{tab:parameters}) while minimising computational resources. We have validated the accuracy of our numerical setup with a suite of small $N$-body simulations (Fig.~\ref{fig:pk_error}) and by presenting a realization of the Euclid comparison project (Fig.~\ref{fig:euclid_pk}). These tests indicate our simulations have an accuracy of 1\% up to $k \sim 5\ihMpc$.    

We have employed our BACCO simulations to predict more than 16,000 nonlinear power spectra at various redshifts and for 800 different cosmologies (Figs.~\ref{fig:emu_coords} and  \ref{fig:emu_data}). These cosmologies span essentially all the currently allowed region of parameter space of $\Lambda$CDM extended to massive neutrinos and dynamical dark energy. Using these results, we built an emulator for the 6 most important principal components of the ratio of the nonlinear power spectrum over the linear expectation (Fig.~\ref{fig:emu_pca}). We show our emulation procedure to be accurate at the $1-2\%$ level over $0 < z < 1.5$ and $10^{-2} < k/(\ihMpc) < 5$ (Figs.~\ref{fig:emu_example} and \ref{fig:emu_test_pk}). Therefore, our accuracy is currently limited by that of cosmology rescaling methods. We compared our predictions against four popular methods to quickly predict the power spectrum in the minimal $\Lambda$CDM scenario (Fig.~\ref{fig:emu_comparison}) and in the presence of massive neutrinos (Fig.~\ref{fig:emu_nu}). 

Since predicting a given cosmology requires an almost negligible amount of CPU time in our BACCO framework, we foresee the accuracy of our emulator to continuously improve as we include more cosmologies in the training set. Extensions to more parameters should also be possible, as, for instance, the number of relativistic degrees of freedom or curvature, can be easily incorporated in cosmology-rescaling methods. Additionally, there are several aspects of such methods that are likely to improve in the future, which should feedback into more accurate predictions and emulated power spectra. 

On the other hand, effects induced by baryons on the shape of the nonlinear mass power spectrum can be of 10-30\% \citep[e.g.][]{Chisari:2019}. Thus, they are much larger than current uncertainties in our emulation, cosmology-rescaling, or even shotnoise. These effects of star formation, gas cooling, and feeback from supermassive black holes are quite uncertain and differ significantly between different hydrodynamical simulations. However, they can be accurately modelled in postprocessing using dark-matter only simulations \citep{S&T2015}. Specifically, \cite{Arico:2020} showed that the effects of 7 different state-of-the-art hydrodynamical simulations could all be modelled to better than 1\% within simple but physically-motivated models. In the future, we will implement such models and extend our matter emulation to simultaneously include cosmological and astrophysical parameters \citep{Arico:2021}.

\section*{Acknowledgments}

The authors acknowledge the support of the ERC-StG number 716151 (BACCO). SC acknowledges the support of the ``Juan de la Cierva Formaci\'on'' fellowship (FJCI-2017-33816). We thank Simon White, Auriel Schneider, Volker Springel for useful discussions. We thank Aurel Schneider also for making public the initial conditions of the EUCLID comparison project, and to Lehman Garrison for providing us with the respective power spectra of the ABACUS, Ramses, PkdGrav, and Gadget simulations. The authors acknowledge the computer resources at MareNostrum and the technical support provided by Barcelona Supercomputing Center (RES-AECT-2019-2-0012).

\section*{Data Availability}

The data underlying this article will be shared on reasonable request to the corresponding author.






\bibliographystyle{mnras}
\bibliography{bibliography} 


\bsp	
\label{lastpage}
\end{document}